# 1. INTRODUCTION

In spite of numerous stunning successes of Quantum Mechanics, its description of the underlying process of measurement [1] is widely considered as inadequate. Bell's inequality [2] and subsequent experimental confirmation [3] had led physicists to accept quantum nonlocality[4] . But the measurement problem in Quantum Mechanics has sharply bared the incompleteness of the present form of the theory in the sense that though a quantum system in a superposed state evolves according to unitary time evolution, prescribed by Schroedinger equation, the measurement outcome chooses a definite state instead of the theoretically expected superposition. While Schroedinger unitarity offers an entangled state for the composite system i.e. the Quantum System (QS) and the measuring device, the measurement process somehow disentangles the state and gives over a definite single outcome. This last process is not contained in the theoretical framework of Quantum Mechanics. While confronting with such an exotic process, Dirac and later von Neumann introduced projection postulates [5] which the theory of Quantum Mechanics per se viewed as ad hoc. The question where Classical Physics and Quantum Mechanics have an interface has never been answered convincingly.

There was, of course, some progress in different directions to resolve the measurement problem, which looks for the actual process underlying the reduction of state vectors:

$$|\psi\rangle = \sum_i c_i |\psi_i\rangle \rightarrow |\psi'\rangle = |\psi_j\rangle$$

where i = 1,2,3,….,and j = 1, or , 2 , or 3,…… The following resolutions carried along with them some visibility in an otherwise opaque problem:

   1) Spontaneous Localization Model [6]

   2) Continuous Spontaneous Localization Model [7]

   3) Decoherence Approach [8]

4) Many Hilbert Space Approach [9]

5) Gravitationally induced wave function collapse [10]

6) Bohmian Approach [11]
7) Many Worlds Interpretation [12] .

All the above methods have provided some insight into the problem, but almost invariably they have raised certain sharp questions whose answers are not yet ready. The main focus is to find a ``collapse`` model, or a way to recover non-unitarity from Schroedinger equation, or its slightly modified forms [6]. The key question is how to recover the non-unitary evolution from the unitarity of standard Quantum Mechanics with the norm conserved. Attempts to resolve the problem have also been made by the "stochastic" seekers of Schroedinger equation [13].

In this paper I have tried to recover a clear-cut non-unitarity between two successive measurements while unitarity rules the roost at the time of measurement. And all these from standard Quantum Mechanics. In section (2) we derived a new wave-function of a free particle by inserting an ansatz in Schroedinger`s equation. Its squared modulus is found to be a probability density wave. ``Wave mechanics`` now acquires transparency with mysterious `$\psi$` waves now generating more tangible probability waves. The wave function can be normalized perfectly in a straightforward way. The dynamics of a QS shows that it evolves non-unitarily between measurements. The chief reason is that all operators representing the observables of the QS are non- Hermitian normal operators defined in rigged Hilbert Space. For example, the difference between the Hamiltonian H and its adjoint is not zero and this metamorphoses the unitary evolution as non-unitary between measurements. Once non-unitary evolution is established it is straightforward to prove that a pure state evolves into a mixed state. The non-Hermitian operators are, however, transformed into Hermitian ones at the measurement point (MP). This restores unitarity at the measurement time. In section (4) we discuss the derivation of Dirac projection postulate . In section (5) we show the strict divide between the Quantum domain and Classical domain. It is surprising that a classical space-time point is created momentarily at the measurement point – which leads us to the fact that Quantum theory is a universal theory valid for both micro and macro systems. In section (6) we derive the general wave function of QS in any external potential and show via Sturm-Liouville problem [14] that these are energy eigenfunctions. Reduction of state vector and derivation of von Neumann's projection postulate are obtained in Section(7). The problem of preferred basis is tackled in section (8). Section (9) reveals the probability field created by a moving QS. Section (10) shows that a QS always travels as a whole particle concealed in complex space-time while behaving as a probability wave in real space-time during this period.

When the QS reaches the MP particle (or device particle) it emerges from an otherwise inaccessible complex space- time to a real classical space- time point. In quantum theory all measurement results are recorded by measuring device at real classical points. Before this measurement, all eigenvalues of the observable of a QS are complex-valued--- which cannot be measured by devices because the QS has not yet reached the apparatus. This unambiguously answers the question of ``possessed properties`` of a QS before measurement------- a central question in EPR paradox.The complex space- time and spin space elevate Quantum Mechanics in ten dimensions where five imaginary dimensions are compactified as these are obviously non-measurable quantities --- thus showing a subtle connection with superstring theories [15]. Another importance of section (10) is that it firmly places the ``parameter`` time as a non-Hermitian observable in Quantum Mechanics. So, space and time are now on an equal footing, as strongly craved in Quantum Field Theory and Quantum Gravity.

In section (11) while discussing quantum cosmology, we examine the no-boundary proposal of Hartle and

Hawking[16] and Quantum Mechanics finds it quite easy to accommodate the smooth entry of the Quantum universe in imaginary time, and it remains concealed there without any breakdown of quantum laws. Quantum Mechanics is still valid there.Stochasticity of position observables has been used to remove the big bang singularity[17]. With the help of Morera`s theorem of complex analysis [18] we have found that the universe may have been created from nothing because the net probability of its entry and emergence from complex space-time is zero. This bears a relation to the idea suggested by Vilenkin [19]. Here we also mention information loss paradox of black holes. S.W. Hawking first suggested this loss [20] but very recently retracted and declared that there is no information loss [21]. Surprisingly,what we found is that, owing to non-unitary evolution, when no measurements are made,mathematics shows a decrease in entropy of an evaporating black hole. This indicates a startling fact : A gain in information ,thus pointing towards order from disorder in a quantum –mechanical universe! But when measurement is made, entropy increases and a genuine loss of information occurs.

In section (12) we answered the question: why probability density can be measured at any spatial point and time although these are not measurement points. In sec (13) we deduced the uncertainty relation in complex space and showed that Heisenberg`s uncertainty relation is its special case. Quantum Mechanics, quite astonishingly, accommodates even complex and negative probabilities. A built-in normalization procedure of all wave functions is discussed in section (14). It may be added that calculations are shown explicitly in order to reach a wider audience.

2. **NORMALIZABLE WAVE FUNCTION OF A FREE PARTICLE AND MEASUREMENT OF ITS OBSERVABLES**

The plane wave representing a free quantum system (QS)

$$\psi(x,t) = A\exp[ikx - i\omega t]$$

when normalized, diverges. The divergence is usually cured by box [22] or Dirac delta function normalization [23]. Sometimes the rigged Hilbert space triplet [24], which includes the extended space $\Omega^\times$, is invoked to yield a convergent integral by restricting the wave number k to only real values. To obtain a perfectly normalizable wave function of a free particle, a generalized wave function ansatz (WA)

$$\psi(x,t) = \sqrt{P(x,t)}\exp[i\varphi(x,t)]$$

is developed from the position probability density $P(x,t) = P_1(x)P_2(t)$. Assuming the splitting

$$\varphi(x,t) = \varphi_1(x) + \varphi_2(t),$$

the ansatz

$$\psi(x,t) = \sqrt{P_1(x)P_2(t)}\exp\{i\varphi_1(x) + i\varphi_2(t)\} \tag{1}$$

is inserted into time-dependent Schroedinger equation for free particle. After separating the real and imaginary parts of both sides, the following two equations are obtained:

$$\hbar\frac{d\varphi_2}{dt} = \left(\frac{\hbar^2}{2m}\right)\left[\frac{1}{2P_1}\frac{d^2P_1}{dx^2} - \frac{1}{4P_1^2}\left(\frac{dP_1}{dx}\right)^2 - \left(\frac{d\varphi_1}{dx}\right)^2\right] = -E \tag{2}$$

and

$$\left(\frac{-1}{P_2}\right)\frac{dP_2}{dt} = \frac{\hbar}{mP_1}\frac{d}{dx}\left(P_1\frac{d\varphi_1}{dx}\right) = -R. \tag{3}$$

E and R are real constants. We assume $\psi(x,t) \neq 0$. The wave number k and frequency $\omega$ are usually defined as

$$k = \left|\frac{d\varphi_1}{dx}\right|$$

and

$$\omega = \left|\frac{d\varphi_2}{dt}\right|.$$

These definitions lead to

$$\frac{d\varphi_1}{dx} = \pm k$$

and

$$\frac{d\varphi_2}{dt} = \pm \omega.$$

The solution of Schroedinger equation is now a four-component wave function

$$\psi(x,t) = \begin{bmatrix} \psi_{+k,+\omega} \\ \psi_{+k,-\omega} \\ \psi_{-k,+\omega} \\ \psi_{-k,-\omega} \end{bmatrix}$$

which is the non-relativistic counterpart of the solution of Dirac equation [25]. $\psi_{+k,-\omega}$ and $\psi_{-k,-\omega}$ ($\psi_{+k,+\omega}$ and $\psi_{-k,+\omega}$) represent positive (negative) energy particles along $\pm x$ directions respectively. (The negative energy solutions of Schroedinger , Klein-Gordon or Dirac equations [26] do not pose any problem. As will be shown below, the negative energy of a particle exists only at the field point or measurement point (MP) when the particle reaches there. At that space-time point, a momentary classical point is created that is ruled by laws of classical physics. The negative energy solutions therefore may be dismissed as unphysical, as is usually done in classical physics. At other space-time points these solutions carry complex energy eigenvalues whose real parts are negative. A complex value is not measurable by a device, as will be discussed below).

For a positive energy particle moving along +x direction the obvious choices (in order to follow convention) are

$$k = \frac{d\varphi_1}{dx}$$

and

$$\omega = -\frac{d\varphi_2}{dt}.$$

Eq. (2) may be rewritten as

$$E = \hbar\omega = \frac{\hbar^2 k^2}{2m} - \frac{\hbar^2}{4mP_1}\left[\frac{d^2 P_1}{dx^2} - \frac{1}{2P_1}\left(\frac{dP_1}{dx}\right)^2\right] \quad (4)$$

while Eq. (3) yields two expressions for position probability density:

$$P^R(x,t) = P_1(x)P_2(t) = \frac{D}{|k|}\exp\left[R\left(t - \frac{x}{v}\right)\right], \text{ for } x - vt \geq 0 \quad (5)$$

and

$$P^L(x,t) = \frac{D}{|k|}\exp\left[R\left(\frac{x}{v} - t\right)\right] \text{ for } x - vt \leq 0. \quad (6)$$

Restrictions involving (x - vt) are imposed in order to satisfy the requirement $P(x,t) \leq 1$. D is a real constant in the above equations. For $x > vt$, $P_1(x)$ decreases with increase in x so that R is positive. Eqs. (5) and (6) show that the probability density is a wave travelling with the same velocity v of the QS. Actually, the probability distribution function, defined by

$$d\pi(x,t) = P(x,t)dx$$

is more fundamental than probability density. P(x,t) at a point is ill-defined without a neighbourhood dx. We shall return to this point later. The momentum and energy eigenvalues of the free particle, derived from Eq. (5) are

$$[\hbar k + (i\hbar R/2v)] \text{ and } [\hbar\omega + (i\hbar R/2)].$$

Schroedinger picked up the classical energy equation to quantize it into its famous form. Klein-Gordon also quantized relativistic energy equation. There is no a priori reason to believe that other classical relations cannot be quantized. At least this is not forbidden. We, therefore, quantize another classical relation involving position of the QS :

To determine position of the QS at a particular time $t_0 < x/v$, the classical equation $s = vt_0$ is quantized:

$$s\psi = \left(\frac{p}{m}\right)t_0\psi = \frac{-i\hbar t_0}{m}\frac{\partial \psi}{\partial x} = \left[vt_0 + (i\hbar R t_0 / 2mv)\right]\psi, \tag{7}$$

where we have used the wave function extracted from Eq. (5) and momentum $p = mv = \hbar k$. We use the same letter for operator and dynamical variable which we hope will not lead to confusion.

At this point we emphasize that there is a clear distinction between the operators s and x. s is the position operator of the QS while x is the corresponding operator for the field point or measurement point (MP).

According to Eq. (7), the position of the QS at any particular time $t_0$ is complex-valued. This unequivocally proves that space coordinates of a QS are complex. We will later find that time is also complex in Quantum Mechanics.

Eq. (4) and (5) together yield

$$E = \hbar\omega = \frac{\hbar^2 k^2}{2m} - \frac{\hbar^2 R^2}{8mv^2}. \tag{8}$$

The last term on the right side of Eq. (8) is quantum potential on which Bohm based his mechanics [27].

Eq. (8) with E=0 implies that a free QS is never at rest. Its least possible momentum is

$$p_{min.} = \pm\left(\frac{\hbar R}{2v}\right). \tag{9}$$

This is a striking result in that a QS having no energy inches forward with this lowest momentum. As shown below, normalization of wave function yields $R = v$. The lowest momentum of Eq.(9) is thus $\pm\frac{\hbar}{2}$, very small indeed!

Eq. (5) and (6) require the condition

$$x = vt \tag{10}$$

as a measurement event because the particle has now arrived at the MP. The probability density is then

$$P(x,t) = (D/|k|) = \text{constant},$$

which implies that $P_2(t) = $ constant. Eq.(3) then derives another important condition for a measurement event :

$$R = 0. \tag{11}$$

Both the conditions Eq. (10) and (11) must be satisfied by a QS at the MP. The quantum system's position at the MP is retrieved from Eq.(7):

$$s = x = vt \tag{12}$$

where the time of arrival at x,

$$t = \frac{x}{v}$$

is the time of measurement. Since both x and t are independent variables, they are constants at the measurement event. The Hamiltonian and momentum operators, when act on the wave functions derived from Eq. (5) and (6)

$$\psi^R(x,t) = \sqrt{D/|k|} \exp\left[\frac{R}{2}\left(t - \frac{x}{v}\right) + ikx - i\omega t\right], \quad \text{for} \quad x \geq vt \tag{13}$$

$$\psi^L(x,t) = \sqrt{D/|k|} \exp\left[\frac{R}{2}\left(\frac{x}{v} - t\right) + ikx - i\omega t\right], \quad \text{or} \quad x \leq vt, \tag{14}$$

are identified as non-Hermitian operators in the entire region except at the MP, where they transform into Hermitian operators on account of Eq. (12). In a forth-coming paper I shall describe the transformation of angular momentum and spin-component operators from Non-Hermiticity to Hermiticity at the MP. The complex energy eigenvalues of the states described by Eqs. (13) and (14) should not be confused with those of decaying and growing states (Gamow vectors). It will be shown in section (10) that a QS lives in complex space-time, and the complex eigenvalues of its observables are generated by the action of operators in that manifold.

Since the QS is found at the MP at $t = \frac{x}{v}$,

$$P\left(x, \frac{x}{v}\right) = 1.$$

Eqs. (5) and (6) yield $|k|$ as the value of D. The free particle wave functions may now be rewritten from Eqs. (13) and (14):

$$\psi^R(x,t) = \exp\left[\frac{R}{2}\left(t - \frac{x}{v}\right) + ikx - i\omega t\right] \text{ with } \pi^R(x,t) = P^R(x,t) = \exp\left[R\left(t - \frac{x}{v}\right)\right], \text{ for } x \geq vt \tag{15}$$

and

$$\psi^L(x,t) = \exp\left[\frac{R}{2}\left(\frac{x}{v} - t\right) + ikx - i\omega t\right] \text{ with } \pi^L(x,t) = P^L(x,t) = \exp\left[R\left(\frac{x}{v} - t\right)\right], \text{ for } x \leq vt. \tag{16}$$

At x = vt,

$$\psi^R(x,t) = \psi^L(x,t) = \psi(x,t) = \exp[ikx - i\omega t]. \tag{17}$$

The advantage of these wave functions, viz. Eq. (15) is that both position [Eq. (7)] and momentum have sharp values albeit complex. Einstein`s dream of a deterministic Quantum theory, fulfilled in a complex manifold, comes as a surprise; but in real space-time an irreducible indeterminism lingers, owing to Heisenberg`s uncertainty relations. As an experimental test of the wave function (15), the probabilities of finding a particle at a large number of adjacent points along x direction may be measured, and these may be plotted against x to find whether $\pi(x,t)$ is an exponentially decreasing function when x is increased and t is kept fixed, provided $x - vt > 0$.

Note that Eq.(15) verifies that $P^R(x,t)$ and $\psi^R(x,t)$ are continuous on the right side while $P^L(x,t)$ and $\psi^L(x,t)$ of Eq.(16) confirm that they are continuous on the left. At the common boundary of $\psi^R$ and $\psi^L$ at the MP, Eq.(17) shows that the wave is a plane wave at the interface where P(x,t) = 1.The latter result unequivocally reveals a particle at the MP while it is still described by a plane wave.This supports the claim of Ghose,Home and Agarwal[27] and its experimental test [28] that wave and particle natures are not mutually exclusive.

When $x \neq vt$, the non-Hermitian operators yield complex eigenvalues from Eqs.(15) and (16). A measuring device cannot register the complex eigenvalues as the QS has not yet reached the MP. The non-Hermitian Hamiltonian H transforms the time –evolution operator

$$U(t,0) = \exp\left[\frac{-iHt}{\hbar}\right]$$

from unitary to non-unitary –and this is obtained from Schroedinger equation! The total probability is, however, conserved:

$$\langle \psi | \psi \rangle = \lim_{\varepsilon \to 0}\left[\int_{vt+\varepsilon}^{\infty} P(x,t)dx\right] = \frac{v}{R}.$$

The total probability is constant, but is not equal to one, when no measurement is made on it.To make it one, requires a measurement and the fact that always a whole particle is obtained at the MP. This occurs because a transition from non-unitarity to unitarity takes place by way of normalization at the MP. Normalization yields R = v. Eqs.(15) and(16) now yield very simple formulas for probability densities of incoming and outgoing quantum systems : $\exp(vt - x)$ for $x \geq vt$, and $\exp(x - vt)$ when $x \leq vt$. From Eq. (15) in Dirac notation

$$H|\psi\rangle = \left[\hbar\omega + \left(\frac{i\hbar R}{2}\right)\right]|\psi\rangle,$$

$$\langle\psi|H^\dagger = \langle\psi|\left[\hbar\omega - \left(\frac{i\hbar R}{2}\right)\right]$$

$$H^\dagger|\psi\rangle = \left[\hbar\omega - \left(\frac{i\hbar R}{2}\right)\right]|\psi\rangle$$

and

$$\langle\psi|H = \left[\hbar\omega + \frac{i\hbar R}{2}\right]\langle\psi|.$$

These equations establish that H is a normal operator. And it is almost ubiquitous in Quantum Mechanics that most operators are non-Hermitian normal operators, as we shall see later. These normal operators are defined in the domains of two non-intersecting rigged Hilbert Spaces for the incoming or outgoing QS.The stochastic position of the MP,discussed in section (13), provides the required non-intersection of the two spaces.

## 3. NON-UNITARY TIME EVOLUTION OF A PURE STATE INTO A MIXED STATE

In this section we will show that the non-unitary time evolution of a pure state results in a mixed state. Consider the following pure state at t = 0:

$$|\psi(0)\rangle = a|\psi_1(0)\rangle + b|\psi_2(0)\rangle + c|\psi_3(0)\rangle \qquad (18)$$

If $U(t,0)$ is the time-evolution operator, the state of the QS at time t is

$$|\psi(t)\rangle = \exp\left(\frac{-iHt}{\hbar}\right)|\psi(0)\rangle = U(t,0)|\psi(0)\rangle \text{ and } \langle\psi(t)| = \langle\psi(0)|\exp\left(\frac{iH^\dagger t}{\hbar}\right) = \langle\psi(0)|U^\dagger(t,0)$$

The initial and final density matrices are respectively

$$\rho_0 = |\psi(0)\rangle\langle\psi(0)| \text{ and } \rho_t = |\psi(t)\rangle\langle\psi(t)| = U\rho_0 U^\dagger.$$

Since $\quad U^\dagger U = \exp\left\{\frac{it}{\hbar}(H^\dagger - H) + \frac{it}{\hbar}[H^\dagger,-H]\right\}$ and $UU^\dagger = \exp\left\{\frac{it}{\hbar}(H^\dagger - H) + \left(\frac{it}{\hbar}\right)[-H,H^\dagger]\right\}.$ (19)

The final result is

$$UU^\dagger = \exp\left\{\left(\frac{-it}{\hbar}\right)(H - H^\dagger)\right\} \qquad (20)$$

because H is a normal operator. H being non-Hermitian, $U^\dagger U = UU^\dagger \neq 1$.
This implies that $U(t,0)$ is a non-unitary normal operator. The non-Hermitian Hamiltonian makes it non-unitary when the QS is left to itself, i.e. no measurement is made on it[24a]. The norm is not conserved during this period of time. This is not at all an unwelcome thing in Quantum Mechanics. It eludes detection as no measurement is possible during this time. This non-unitarity in the intermediate period may allow creation and destruction of other particles between two measurements. This is very much needed in any relativistic version of quantum theory. But what is striking is that it is present even in the non-relativistic regime. The relativistic Dirac equation had already raised this question whether it is a single –particle theory. And non-relativistic Quantum Mechanics now asks the same question whether its four component wave function and virtual particles are parts of Schroedinger equation. But this problem (if it is a problem at all) is resolved when it is found that unitarity is retrieved at the measurement point. Indeed we have shown it below at the end of this section.

We now switch over from an individual QS to an ensemble of quantum systems in the same state to show that a pure case non-unitarily evolves into a mixed state at the MP. To show the loss of coherence terms in the post-measurement density matrix, let the initial pure state be

$$\rho_0 = |\psi_0\rangle\langle\psi_0|$$

at t = 0. The pure state of the QS may be written as
$$|\psi_0\rangle = a|\psi_{10}\rangle + b|\psi_{20}\rangle + c|\psi_{30}\rangle. \qquad (21)$$

During non-unitary time evolution $|\psi_0\rangle$ becomes

$$|\psi_t\rangle = \left\{\exp\left(\frac{-iHt}{\hbar}\right)\right\}|\psi_0\rangle = a|\psi_{1t}\rangle + b|\psi_{2t}\rangle + c|\psi_{3t}\rangle. \quad (22)$$

The quantum systems are randomly situated at any of the wave peaks of $|\psi_{1t}|^2, |\psi_{2t}|^2$ and $|\psi_{3t}|^2$ and so $\psi_{1t}, \psi_{2t}$ and $\psi_{3t}$ randomly transform into plane waves at the MP. In any single measurement a QS is at any of the wave peaks of $|\psi_{1t}|^2, |\psi_{2t}|^2$ or $|\psi_{3t}|^2$. We confirm the validity of this statement in the following way: Which of the particular peaks of these waves will reach the MP first, would be decided by the velocities of the respective probability waves. The position operator of the quantum system acts on, say,

$$\left[\psi_{1t}^{R_1}(x,t)\right]_{x=v_1 t} = \left[\exp\left\{\frac{R_1}{2}\left(t - \frac{x}{v_1}\right) + ik_1 x - i\omega_1 t\right\}\right]_{x=v_1 t} = \exp[ik_1 x - i\omega_1 t] \quad (23)$$

Note that position is measurable only when the wave peak reaches x = vt.

Therefore, $\quad s\psi_{1t}^{R_1}(x,t) = \frac{-i\hbar t_1}{m}\frac{\partial}{\partial x}\left[\exp\{ik_1 x - i\omega_1 t\}\right] = v_1 t_1 \psi_{1t}^{R_1}(x,t),$

so that $\quad\quad\quad\quad\quad\quad\quad \langle\psi_{1t}|s|\psi_{1t}\rangle = v_1 t_1$

From Eq.(22),
$$\langle\psi_t|s|\psi_t\rangle = |a|^2\langle\psi_{1t}|s|\psi_{1t}\rangle + |b|^2\langle\psi_{2t}|s|\psi_{2t}\rangle + |c|^2\langle\psi_{3t}|s|\psi_{3t}\rangle$$

$$= |a|^2(v_1 t_1) + |b|^2(v_2 t_2) + |c|^2(v_3 t_3).$$

The possible positions of the QS are at the points $v_1 t_1$ or $v_2 t_2$ or $v_3 t_3$ with respective probabilities $|a|^2, |b|^2$ and $|c|^2$. Without loss of generality, let $\psi_{1t}$ reaches the MP(at x) first so that it becomes a plane wave. If and only if the QS is at $v_1 t_1 = x$ then

$$\langle x|\psi_{1t}\rangle = \exp[ik_1 x - i\omega_1 t]$$

which implies
$$\langle x|\psi_{1t}\rangle\langle\psi_{1t}|x\rangle = 1 = \langle x|\mathbf{1}|x\rangle$$

Therefore,
$$\rho_{1t} = 1. \quad (24)$$

The resolution of the identity operator [30] yields

$$\rho_{1t} + \rho_{2t} + \rho_{3t} = 1$$

where $\rho_{2t}$ and $\rho_{3t}$ are respectively $|\psi_{2t}\rangle\langle\psi_{2t}|$ and $|\psi_{3t}\rangle\langle\psi_{3t}|$. Eq. (24) establishes that

$$\rho_{2t} + \rho_{3t} = 0.$$

Since $\rho_{2t}$ and $\rho_{3t}$ are positive semi-definite

$$\rho_{2t} = \rho_{3t} = 0. \tag{25}$$

Eq. (22) gives

$$\rho_t = |\psi_t\rangle\langle\psi_t| = |a|^2 \rho_{1t} + |b|^2 \rho_{2t} + |c|^2 \rho_{3t}$$
$$+ \rho_{1t}\rho_t\rho_{2t} + \rho_{1t}\rho_t\rho_{3t} + \rho_{2t}\rho_t\rho_{1t} + \rho_{2t}\rho_t\rho_{3t} + \rho_{3t}\rho_t\rho_{1t}$$
$$+ \rho_{3t}\rho_t\rho_{2t}.$$

Eq. (25) causes the loss of coherence terms in $\rho_t$ which reduces to

$$\rho'_1 = |a|^2 \rho_{1t} \tag{26}$$

This happens for $|a|^2$ times. Similarly, the states $|\psi_{2t}\rangle$ and $|\psi_{3t}\rangle$ transform into plane waves at the MP and yield the subensembles of density matrices $|b|^2 \rho_{2t}$ and $|c|^2 \rho_{3t}$ respectively. The total density matrix equals the sum of the density matrices of these subensembles:

$$\rho_t \to \rho'_t = \rho'_1 + \rho'_2 + \rho'_3 = |a|^2 \rho_{1t} + |b|^2 \rho_{2t} + |c|^2 \rho_{3t} \tag{27}$$

Now, $$\rho'^2_t = |a|^4 \rho_{1t} + |b|^4 \rho_{2t} + |c|^4 \rho_{3t}$$

Hence, $$\rho'^2_t \neq \rho'_t$$

and $$\text{trace}\left(\rho'^2_t\right) = |a|^4 + |b|^4 + |c|^4 < 1.$$

$\rho'_t$ is therefore a mixed state. As we have stated earlier, one more spectacular thing happens at the MP. While $\rho'_t$ becomes a mixed state, the non-Hermitian H transforms into a Hermitian H ( each time a reduction occurs ) on account of formation of the plane wave at the MP. Eq. (20) retrieves the unitarity of $U(t,0)$ at the MP:

$$U^\dagger U = UU^\dagger = 1.$$

## 4. DERIVATION OF DIRAC PROJECTION POSTULATE

We now derive Dirac projection postulate from the mechanism of measurement on an individual QS. We have seen in the previous section that a QS will be randomly distributed at the definite positions $v_1 t_1, v_2 t_2$ or $v_3 t_3$ (at the respective probability wave peaks) for $|a|^2, |b|^2$ or $|c|^2$ times respectively. This does not contradict Kochen-Specker theorem [30] because a QS cannot have a pre- measurement (real-valued) definite outcome for any of its observables.

Let any of the component wave functions, e.g. $a\psi_1$ reaches the MP first. The presence of the QS at $v_1 t_1$ in any particular measurement is totally unpredictable. When a QS is actually at $x = v_1 t_1$ in a particular measurement, the reference frame of the QS as its origin coincides with the reference frame of the MP particle. The wave function of the QS is changed to a plane wave of Eq. (23) and probability density of the wave is one. Hence all observables are obtained as real eigenvalues corresponding to the plane wave and a whole particle, with $P\left(x, \frac{x}{v_1}\right) = 1$, emerges at the MP. In other words, a momentary classical point opens a window of real space-time at the MP at $t_1 = \frac{x}{v_1}$. If no permanent record of the QS is made at the MP, the plane wave enters the zone with a wave function represented by Eq.(16) with the condition $x \leq vt$.

The `When` and `Where` of the measurement are respectively $t_1$ and $x(= v_1 t_1)$. In a particular measurement the MP at x (and automatically, at time t) is fixed by the observer who places the required ``device particle `` at that space-time point. This constitutes the selection of preferred basis of measurement. This is supported by Copenhagen Interpretation [31]. The ``How`` of the measurement process is described by transformation of non-Hermitian operators into Hermitian ones at the measurement point.

Once the QS is found at $x = v_1 t_1$ it will not show up for the remaining $\left(1 - |a|^2\right)$ times. In such cases, the reference frame of the QS will not be available. The state $\psi_1$ will not transform into a plane wave because of the absence of coincidence of the two frames of QS and MP. The criterion $x = v_1 t_1$ is not fulfilled.

Consider now the measurement process on an individual QS. According to relative frequency interpretation [32], if $|a|^2 = \frac{1}{2}$ in a run of 100 measurements, then 50 quantum particles will be found at the MP in the plane wave state. If one wants to find out whether there is a particle at the MP in any single measurement he must fall back on the much-debated topic of subjective probability [33].
Since the MP is a classical point at the measurement event and the QS leaves the quantum domain to enter into classical domain, quantum probability switches over to classical probability, where in the latter case the probability is either zero or one. In order to clarify the last point we need prior probability and posterior probability in the Bayesian approach [34]. Bayesians claim that it is always possible to determine such a prior distribution to specify subjective probability. L.J.Savage [35] was able to develop a theory of subjective probability from a set of postulates. Before the probability wave $|\psi_1|^2$ reaches the MP, the quantum probability is $|a|^2 < 1$. Empirically, a whole particle is found at the MP. An abrupt change therefore occurs as the QS exits the quantum domain and reaches the momentary classical point at the MP. The abrupt changes are :

> i) The state $a|\psi_1\rangle$ changes to $|\psi_1'\rangle$ at the MP, which represents a plane wave. The observables act on $\psi^R(x,t)$, [Eq.(15)], which is a continuous function and where the operations are carried out at $x = v_1 t_1$ .

ii) The equation representing the pure state $|\psi\rangle = a|\psi_1\rangle + b|\psi_2\rangle + c|\psi_3\rangle$

changes to $|\psi'\rangle = |\psi_1'\rangle + b'|\psi_2'\rangle + c'|\psi_3'\rangle$

because the amplitudes $a = \langle\psi_1|\psi\rangle, b = \langle\psi_2|\psi\rangle$ and $c = \langle\psi_3|\psi\rangle$

change to $a' = \langle\psi_1'|\psi'\rangle = 1, b' = \langle\psi_2'|\psi'\rangle$ and $c' = \langle\psi_3'|\psi'\rangle$.

iii) Relative frequency interpretation of probability, requiring a large number of experiments, is abandoned, and replaced by an interpretation that ascribes probability to single events. This is needed if one insists on showing the reduction of state vector in an individual measurement[36]. Now classical probability takes over from quantum probability at the MP.

Before the measurement the total probability was

$$|a|^2 + |b|^2 + |c|^2 = 1$$

and at the time of measurement it is

$$1 + |b'|^2 + |c'|^2 = 1$$

yielding $b' = c' = 0$. This clearly indicates a reduction of state vectors. At the MP the unitarity of time evolution operator is restored, that is, probability is conserved at this point. Since probability cannot be destroyed, at the time of reduction the probability contents $|b|^2$ and $|c|^2$ are instantaneously transported to the probability $|a|^2$ at the MP to add up to one. At the end of measurement Eq.(21) becomes

$$|\psi'_{t_1}\rangle = |\psi'_{1t_1}\rangle \tag{32}$$

where $t_1 = \dfrac{x}{v_1}$. Thus a definite outcome in an individual measurement has been ensured. That means Dirac projection postulate has been derived, as pointed out by Ghirardi [36]. If no QS has been found at $v_1t_1$, it would surely be found at either $v_2t_2$ or $v_3t_3$. In the latter cases the outcomes will be $|\psi'_{2t_2}\rangle$ or $|\psi'_{3t_3}\rangle$ with the respective probabilities $|b|^2$ and $|c|^2$. This implies that the distinguishibility among the three outcomes is realized. The von Neumann projection postulate is discussed in Sec(7).

## 5. MOMENTARY CLASSICAL SPACE-TIME POINT IN QUANTUM MECHANICS

The classical position coordinate of the QS at s = x = vt is characterized by a real number. When $x \neq vt$ the position coordinate of the QS is a complex number. The measurement points of Quantum Mechanics characterized by x = vt are the position coordinates of a moving classical particle. This analogy labels classical physics as "measurement physics". Classical physics is valid only for a momentary merger of the two reference frames containing respectively the QS and the MP particles. Classical physics is thus reduced to a special case of Quantum Mechanics.

To prove that when $P(x,t) = 1$, a classical space-time point emerges, we consider the quantum-mechanical counterpart of Hamilton-Jacobi equation [29]:

$$\frac{\partial s}{\partial t} + \frac{1}{2m}\left(\frac{\partial s}{\partial x}\right)^2 - \frac{i\hbar}{2m}\frac{\partial^2 s}{\partial x^2} + V(x) = 0 \tag{33}$$

where $s(x,t)$ is defined by

$$\psi(x,t) = \exp\left[\frac{is}{\hbar}\right] = \exp[ik_1 x - i\omega_1 t]$$

so that $s(x,t) = \hbar(k_1 x - \omega_1 t)$ and $\frac{\partial^2 s}{\partial x^2} = 0$. Eq.(33) is now the quantum-mechanical counterpart of Hamilton – Jacobi equation for the Hamilton's principal function $s(x,t)$ of classical Mechanics. Classical physics is inducted into Quantum Mechanics because the quantum potential in Eq.(4), rewritten as

$$\frac{-\hbar^2}{4mP_1}\left[\frac{\partial^2 P}{\partial x^2} - \frac{1}{2P}\left(\frac{\partial P}{\partial x}\right)^2\right]$$

[where P(x,t) is given by Eq.(15)] vanishes at the MP at x = vt where $P^R(x,t) = 1$. Quantum Mechanics subsumes Classical Physics, and the former thus decidedly becomes a universal theory ruling both macro and micro-world.The relation x = vt for a classical particle now acquires a deeper meaning in that Classical Physics tacitly assumes that Quantum-mechanical measurements are always being made at all values of x along the classical particle's trajectory. This might label Classical Physics as`` Quantum Measurement Physics``.A question might arise that if this universal Quantum Mechanics is embedded in complex space-time, how we are able to see a moving classical object. The answer lies in swapping ``seeing``with ``measuring position eigenvalues``.Classical things live in the real- dimensional parts of the complex-dimensional space-time of Quantum Mechanics. And this happens to be the right place to seek an answer to Einstein's famous question to Abraham Pais on the existence of the moon when one is not looking at it. When one looks, quantum measurement of position operator of the photon (carrying the configuration-information of the moon) from the moon is made and he gets a real value enabling him to see the moon in real space-time.When he is not looking, the direction of motion of a photon or photons will not coincide with the line of sight at the end of which is the MP particle in the eye..The probability wave peak of the photon will not be able to reach the MP, and consequently a complex eigenvalue of the photon's position will be registered indicating that a measurement has not taken place.This photon in complex space-time will,through one-to-one correspondence, carry the information that the moon is in complex space-time. The moon, therefore, very much exists when one is not looking at it. It then resides in the unobservable complex space-time. The quantum dynamics of the non-free states discussed in Sec.(6) are not motionless and

$$\frac{\hbar}{m}\frac{d\phi_1}{dx} = v(x,t) = \frac{1}{m}\frac{\partial s}{\partial x}$$

is the particle velocity at x. The velocity v(x) refers to particle velocity at the classical point x, and therefore the uncertainty relation of position-momentum cannot prevent its use.

## 6. THE MOST GENERAL ENERGY EIGENFUNCTION OF A QUANTUM SYSTEM IN A POTENTIAL

In this section I derive the most general wave function of a quantum system by inserting the WA of Eq. (1) in the time- dependent Schroedinger's equation. The potential V(x) is time-independent. Two equations are obtained by separating the real and imaginary parts of both sides:

$$\hbar \frac{d\phi_2}{dt} = \frac{\hbar^2}{2m}\left[\frac{1}{2P_1}\frac{d^2P_1}{dx^2} - \frac{1}{4P_1^2}\left(\frac{dP_1}{dx}\right)^2 - \left(\frac{d\phi_1}{dx}\right)^2\right] - V(x) = -E \tag{34}$$

and

$$-\frac{1}{P_2}\frac{dP_2}{dt} = \frac{\hbar}{mP_1}\frac{d}{dx}\left(P_1\frac{d\phi_1}{dx}\right) = -R \tag{35}$$

where E and R are real constants. As in case of free particles we choose

$$\frac{d\phi_1}{dx} = k(x) \quad \text{and} \quad \frac{d\phi_2}{dt} = -\omega$$

for a QS travelling with positive energy along + x direction. Eq. (35) yields the following:

$$\left.\begin{array}{c} P_2(t) = A\exp(Rt) \\ \\ \text{and} \quad P_1(x) = \dfrac{B}{|K(x)|}\exp\left[-R\int\dfrac{dx}{v(x)}\right] \end{array}\right\} \tag{36}$$

Combining these two expressions, the position probability density takes the forms

$$P(x,t) = \frac{D}{|K(x)|}\exp\left[R\left(t - \int\frac{dx}{v(x)}\right)\right], \text{ for } t \leq \int\frac{dx}{v(x)} \tag{37}$$

and

$$P(x,t) = \frac{D}{|K(x)|}\exp\left[R\left(\int\frac{dx}{v(x)} - t\right)\right], \text{ for } t \geq \int\frac{dx}{v(x)} \tag{38}$$

where D=AB, a real constant. Eqs.(37) and (38) describe position probability density of the QS as an inhomogeneous probability wave travelling with the same variable velocity v(x) of the QS. We reiterate that the point s = x belongs to the classical regime as shown at the end of Sec.(5). The corresponding wave functions of the QS are

$$\psi(x,t) = \sqrt{\frac{D}{k(x)}} \exp\left[\frac{R}{2}\left(t - \int\frac{dx}{v(x)}\right) + i\int k(x)dx - i\omega t\right] \quad \text{for } t \leq \int\frac{dx}{v(x)} \tag{39}$$

and

$$\psi(x,t) = \sqrt{\frac{D}{k(x)}} \exp\left[\frac{R}{2}\left(\int\frac{dx}{v(x)} - t\right) + i\int k(x)dx - i\omega t\right] \quad \text{for } t \geq \int\frac{dx}{v(x)} \tag{40}$$

A straightforward calculation from Eq.(35) and any of the equations (39) or (40) derives the meaning of $\psi(x,t)$: it is a function of probability density, f [P(x,t)]. Making use of Eq.(37) the two results.

$$\frac{\partial P}{\partial t} = RP \quad \text{and} \quad \frac{\partial}{\partial x}[P(x,t)v(x)] = -RP$$

combine to yield the continuity equation $\quad \dfrac{\partial P}{\partial t} + \dfrac{\partial}{\partial x}[P(x,t)v(x)] = 0$, \tag{41}

where v(x) is the velocity of the classical particle at the point x of classical domain.

For the incoming wave towards the MP[ Fig (1)] , P(x,t) is described by

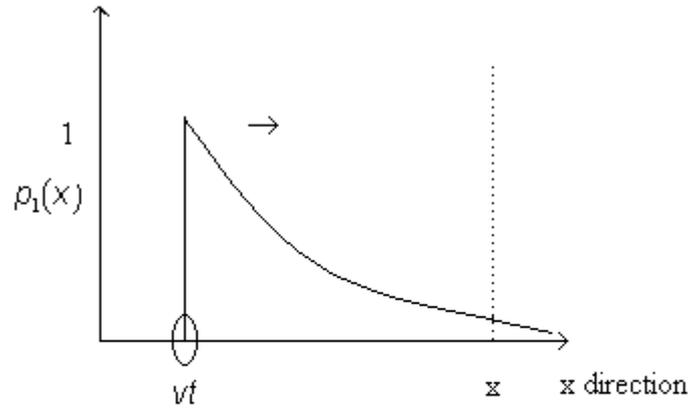

Figure (1)
The probability wave p(x,t) of a quantum system
approaching the measurement point x when x > vt.
The particle is at vt at time t.

Eq.(37) in the space interval extending from the position of the wave peak to infinity. When the QS reaches the MP the probability wave is still described by Eq. (37) in the interval $\left[x = \int vdt, \infty\right)$ [Fig. (2)]:

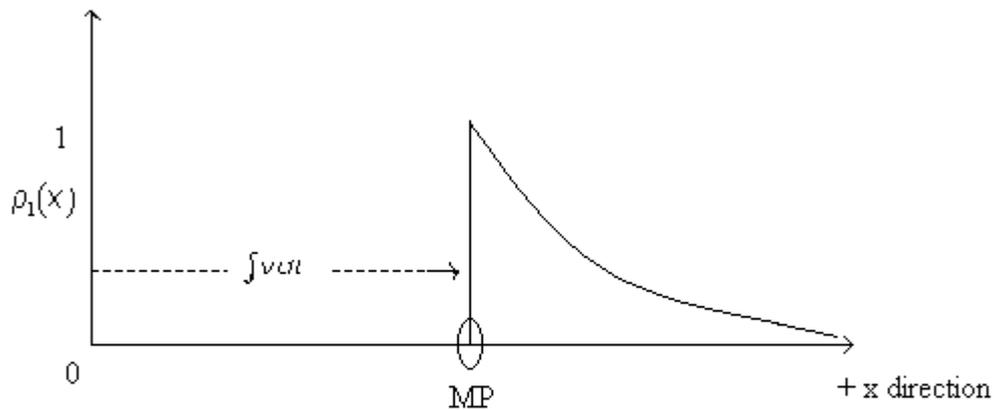

Figure(2): The QS is at the MP at $x = \int v dt$ and the the probability wave retains its previous form.

When the QS crosses the MP the tail of probability wave in Fig (2) instantaneously sweeps out the entire space from $+\infty$ to $-\infty$, and the wave form changes as if reflected by the vertical line at MP. It is now described by Eq.(38) in the space interval $\left(-\infty, \int v dt\right]$, [Fig.(3)]:

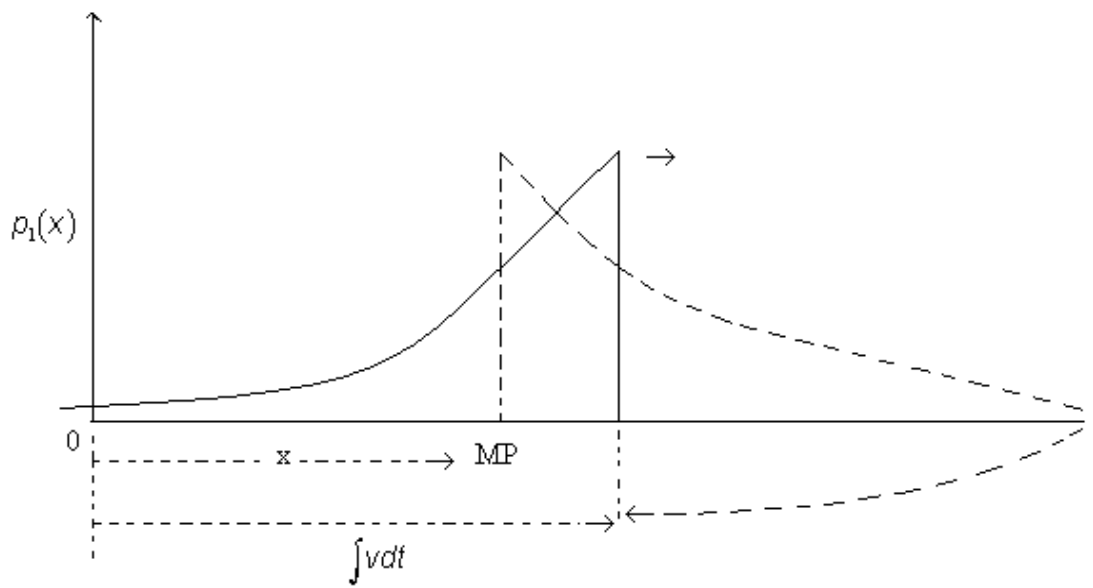

Figure(3): The probability wave of the QS changes its form from the dotted wave to the continuous line form sweeping out the entire space instantaneously in a 180degrees turn-around from $+\infty$ to $-\infty$.

There is an element of non-locality in this type of instantaneous sweeping. But the probability wave is not a physical wave It is a mathematical wave which is travelling in a probability space. It is not a signal in physical space. The probability density wave P(x,t) is constructed by repeating the measurement on identical copies of the QS . This mathematical entity  P(x,t) in physical space need not respect the special relativistic  velocity limit of a physical signal. This fact demystifies the quantum non-locality what Einstein referred to as "spooky action at a distance."

When the QS reaches the MP at $t = \int \frac{dx}{v(x)}$, Eqs.(37) and Eqs.(38) yield

$$P(x,t) = \frac{D}{|k(x)|}. \tag{42}$$

$$\therefore P_2(t) = \text{constant},$$

which together with Eq.(3) yields the important result

$$R = 0 \tag{43}$$

at the MP. The QS co-moves with the peak of the probability wave. Therefore, at the MP,

$$P(x,t) = 1$$

and according to Eq.(42)

$$D = |k(x)| = k, \text{ a real constant} \tag{44}$$

at the MP . So at the MP , the QS in a time-independent potential behaves like a free particle ! At other points in space the wave functions are obtained from Eqs.(39) and (40):

$$\psi(x,t) = \left|\frac{k}{k(x)}\right|^{\frac{1}{2}} \exp\left[\frac{R}{2}\left(t - \int \frac{dx}{v(x)}\right) + i\int k dx - i\omega t\right], \text{ for } t \leq \int \frac{dx}{v(x)} \tag{45}$$

and

$$\psi(x,t) = \left|\frac{k}{k(x)}\right|^{\frac{1}{2}} \exp\left[\frac{R}{2}\left(\int \frac{dx}{v(x)} - t\right) + i\int k dx - i\omega t\right], \text{ for } t \geq \int \frac{dx}{v(x)} \tag{46}$$

From these wave functions one finds that non-Hermitian observables transform into Hermitian ones with real eigenvalues at the MP owing to the relation (44).

To derive eigenfunctions of a QS in an arbitrary time-independent potential V(x), I consider Eq.(34) with the input

$$P_1(x) = (R_1(x))^2.$$

A little rearrangement converts the equation as a Sturm-Liouville problem [14]

$$\frac{\hbar^2}{2m}\frac{d^2 R_1}{dx^2} - \left[\frac{\hbar^2 k^2(x)}{2m} + V(x)\right]R_1 + \hbar\omega R_1 = 0 \tag{47}$$

where $\hbar\omega = E$ is the Sturm-Liouville constant. The boundary conditions are

$$R_1(\infty) = 0 \quad \text{and} \quad \left(\frac{dR_1}{dx}\right)_{x=x_0} = 0$$

where $x_0 = \lim_{\varepsilon \to 0}\left[\varepsilon + \int vdt\right]$. The point $x_0$ is arbitrarily close to the MP. A complete set of energy eigenfunctions (without the time and phase terms) $R_n(x)$ may now be obtained. Replacing $P_1(x)$ of the Eq.(37) or (38) by $R_1^2(x)$ and then eliminating $R_1(x)$ to calculate $k(x)$ may seem a viable route to obtain $R_1(x)$; but even for a harmonic oscillator potential the resulting nonlinear differential equation is quite intractable. Non-conservation of energy is a hallmark of all "collapse models", e.g Ghirardi, Rimini, Weber, Pearle theory[7]. But it is not relevant in this discussion. The complex eigenvalue of energy of a QS is not measurable by a device. The real part of this energy and the system's real valued energy at the MP are the same and this therefore conserves the real part of energy. But the difference of energy eigenvalues before and after a measurement is an imaginary number and therefore eliminates the relevance of the issue. One of the frequently asked questions particularly in EPR paradox[37] is whether a QS possesses a property or value of a dynamical variable before a measurement. It has already been shown that a QS has complex-valued pre-measurement results of its property, but it cannot be measured as the QS has not yet reached the MP. Thus a QS possesses a definite (complex-valued) property before a measurement, but this is not amenable to measurement.

Note that when a QS in a potential reaches the MP at $t = \int \frac{dx}{v(x)}$, Eqs.(43), (44) and (45) or (46) show it behaves as a free QS at the MP!

$$\left[\psi(x,t)\right]_{t=\int \frac{dx}{v(x)}} = \exp[ikx - i\omega t] \tag{48}$$

## 7. REDUCTION OF STATE VECTOR AND DERIVATION OF von NEUMANN`S PROJECTION POSTULATE

Sec.(5) showed that classical physics is a special case of Quantum Mechanics. A wave function may, therefore, be assigned to the macroscopic measuring device. The initial state of the composite system(QS plus device) at $t = 0$ is

$$\psi(0) = [a\psi_1(0) + b\psi_2(0)]\phi_0 \tag{49}$$

where $\psi_1$ and $\psi_2$ refer to two mutually orthogonal eigenstates of the observable $\hat{A}$ of the QS and $\phi_0$ is the pointer state of the device at $t = 0$. In order to know $\psi_0$ an initial measurement has to be made on the composite system at $t = 0$ when the time evolution operator is unitary. Once unitarity starts Eq.(49) transforms to the famous entangled state[44]:

$$\psi(x,t) = a\psi_1(x,t)\phi_1(x,t) + b\psi_2(x,t)\phi_2(x,t). \tag{50}$$

$\psi_1$ and $\psi_2$ are the most general eigenfunctions of the operator $\hat{A}$ of the QS where $t > 0$. $\phi_1$ and $\phi_2$ are mutually orthogonal and macroscopically distinguishable pointer states of the observable B of the device. If the pointer contains N quantum systems then

$$\phi_1 = \phi_1(q_1, q_2, q_3, \ldots\ldots, q_N, t)$$

represents the total normalized wave function of the pointer in an eigenstate. Both $\psi_1$ and $\varphi_1$ satisfy Schroedinger equation

and
$$\left. \begin{array}{l} \dfrac{-\hbar^2}{2m} \dfrac{\partial^2 \psi_1}{\partial x^2} + V_1(x)\psi_1 = i\hbar \dfrac{\partial \psi_1}{\partial t} \\[6pt] \dfrac{-\hbar^2}{2m} \dfrac{\partial^2 \phi_1}{\partial x^2} + V_2(x)\phi_1 = i\hbar \dfrac{\partial \phi_1}{\partial t} \end{array} \right\} \qquad (51)$$

Let $\theta_1(x,t) = \psi_1 \phi_1$, $\theta_2(x,t) = \psi_2 \phi_2$ and $V(x) = V_1(x) + V_2(x)$.

Then
$$\begin{aligned} \dfrac{-\hbar^2}{2m} \dfrac{\partial^2 \theta_1}{\partial x^2} + V(x)\theta_1 &= \dfrac{-\hbar^2}{2m} \psi_1 \dfrac{\partial^2 \phi_1}{\partial x^2} + \dfrac{-\hbar^2}{2m} \phi_1 \dfrac{\partial^2 \psi_1}{\partial x^2} + V_1(x)\psi_1 \phi_1 + V_2(x)\psi_1 \phi_1 \\ &= i\hbar \psi_1 \dfrac{\partial \phi_1}{\partial t} + i\hbar \phi_1 \dfrac{\partial \psi_1}{\partial t} \\ &= i\hbar \dfrac{\partial \theta_1}{\partial t}. \end{aligned}$$

Therefore $\theta_1(x,t)$ (and similarly $\theta_2(x,t)$) is a valid wave function of the composite system (QS + pointer). The device pointer and the QS are separate objects with some distance between them. Nor are they identical systems. Still Quantum Mechanics assigns a wave function of the composite system obeying Schroedinger`s equation. Hence there is no a priori reason why the entire universe should not have a wave function ! In fact this supports Hartle and Hawking`s idea of wave function of the universe [16]. In state vector language the vectors $|\psi_1\rangle$ and $|\psi_2\rangle$ belong to the complex rigged Hillbert space $H_1$. Similarly $|\varphi_1\rangle$ and $|\varphi_2\rangle$ belonging to $H_2$ produce the vectors

and
$$\begin{aligned} |\theta_1\rangle &= |\psi_1\rangle_{H_1} \otimes |\phi_2\rangle_{H_2} \\ |\theta_2\rangle &= |\psi_2\rangle_{H_1} \otimes |\phi_2\rangle_{H_2} \end{aligned} \qquad (52)$$

in the tensor product space $H_1 \times H_2$. This product space is a mathematical space where reductions of the state vectors take place.

Note that
$$\begin{aligned} (A \otimes B)|\theta_1\rangle &= (A|\psi_1\rangle) \otimes (B|\varphi_1\rangle) \\ &= \lambda_1 |\psi_1\rangle \otimes \eta_1 |\phi_1\rangle \\ &= \lambda_1 \eta_1 |\psi_1\rangle \otimes |\phi_1\rangle \\ &= \lambda_1 \eta_1 |\theta_1\rangle \end{aligned}$$

where $\lambda_1$ and $\eta_1$ are eigenvalues corresponding to operators $\hat{A}$ and $\hat{B}$ respectively having eigenvectors $|\psi_1\rangle$ and $|\varphi_1\rangle$. Eqs.(52) therefore help identify the mutual orthogonality of state vectors $|\varphi_1\rangle$ and $|\varphi_2\rangle$ of the composite system:

$$\langle \theta_1 | \theta_2 \rangle = \left( \langle \psi_1|_{H_1} \otimes \langle \phi_1|_{H_2} \right) \left( |\psi_2\rangle_{H_1} \otimes |\phi_2\rangle_{H_2} \right)$$

$$= \langle \psi_1 | \psi_2 \rangle_{H_1} \langle \phi_1 | \phi_2 \rangle_{H_2}$$
$$= 0$$

Returning to wave functions Eq.(50) may be rewritten as
$$\psi(x,t) = a\theta_1 + b\theta_2 \tag{53}$$

which is identical to Eq.(31) except that the probability waves of $|\theta_1|^2$ and $|\theta_2|^2$ are inhomogeneous waves of variable velocity of the composite system propagating in a mathematical space (Tensor product space in case of vectors). In real space-time the QS coupled to an apparatus is synonymous with probability waves except at the MP. The wave functions $\theta_1$ and $\theta_2$ may be written explicitly from Eq.(45):

$$\theta_j(x,t) = \left|\frac{k_j}{k_j(x)}\right| \exp\left[\frac{1}{2}R_j\left(t - \int \frac{dx}{v_j(x)}\right) + i\int k_j(x)dx - i\omega_j t\right] \tag{54}$$

where $j = 1,2$ and $t \leq \int \frac{dx}{v_j(x)}$. When the peak of $|\theta_1|^2$ reaches the MP first, then, from Eq.(44), Eq.(54) reduces to a plain wave
$$\theta_1'(x,t) = \exp[ik_1 x - i\omega_1 t] \tag{55}$$

and $P_1(x,t) = |\theta_1'|^2 = 1$ at the MP. The wave described by Eq.(55) accompany some abrupt changes in the probability amplitudes viz., from a and b to $a' = 1$ and $b' = 0$ respectively. The argument now follows the derivation of Dirac Projection postulate in Sec.(4). Finally it is observed that Eq.(53) undergoes a reduction of state vector or wave function
$$\psi'(x,t) = \theta_1'(x,t)$$
so that from Eq.(50),
$$\psi'(x,t) = \psi_1' \phi_1' \tag{56}$$

where $\psi_1'$ and $\varphi_1'$ are plane waves at the classical point or MP. This happens for $|a|^2$ times. The composite system will be in $\psi_2' \varphi_2'$ for $|b|^2$ times. In Eq.(56) a definite outcome (QS in $\psi_1'$ and pointer in $\varphi_1'$) has been realized for a composite system in an individual measurement event. Eq. (56), showing a reduction of state implies that von Neumann projection postulate has been derived. The irreversibility of the reduction stems from the fact that no time evolution operator can resurrect the pure state Eq. (53) from the post-measurement null state $\theta_2(x,t) = 0$.

One can include the coupling of other devices and environment to the composite system's state, Eq. (50), but that does not change the conclusion because the net initial state may always be described by a pure state of the form of Eq. (53). Consequently von Neumann's chain loses its significance.

## 8. THE PROBLEM OF PREFERRED BASIS

The problem of preferred basis [30] occurs when two ensembles $E$ and $\tilde{E}$ described by the same density matrix are compared:

$$\rho = \frac{1}{2}|\phi_1\rangle\langle\phi_1| + \frac{1}{2}|\phi_2\rangle\langle\phi_2| \tag{57}$$

$$= \frac{1}{2}|A_+\rangle\langle A_+| + \frac{1}{2}|A_-\rangle\langle A_-| \tag{58}$$

where $|\phi_1\rangle = |\uparrow_z\rangle \otimes |h\rangle$ and $|\phi_2\rangle = |\downarrow_z\rangle \otimes |t\rangle$.

The kets $|h\rangle$ and $|t\rangle$ describe pointer states of measuring device:
$$|h\rangle = |\text{ pointer here }\rangle$$
$$|t\rangle = |\text{ pointer there }\rangle$$
and $|\uparrow_z\rangle$ and $|\downarrow_z\rangle$ are spin-up and spin-down states of a QS along z direction.

$|A_+\rangle$ and $|A_-\rangle$ are described by $|A_+\rangle = \frac{1}{\sqrt{2}}[|\phi_1\rangle + |\phi_2\rangle]$ and $|A_-\rangle = \frac{1}{\sqrt{2}}[|\phi_1\rangle - |\phi_2\rangle]$. (59)

Ensemble E contains quantum systems each of which is in state $|\phi_1\rangle$ or $|\phi_2\rangle$. But the quantum systems in the ensemble $\tilde{E}$ are either in state $|A_+\rangle$ or in state $|A_-\rangle$, both of which are in the entangled states

$$|A_+\rangle = \frac{1}{\sqrt{2}}[|\uparrow_z\rangle \otimes |h\rangle + |\downarrow_z\rangle \otimes |t\rangle] \tag{60}$$

and
$$|A_-\rangle = \frac{1}{\sqrt{2}}[|\uparrow_z\rangle \otimes |h\rangle - |\downarrow_z\rangle \otimes |t\rangle]. \tag{61}$$

$|A_+\rangle$ or $|A_-\rangle$ is the state of the composite system (QS plus pointer) in which the QS interacted with the device in one of the its eigenstates of $\sigma_x$. When $\sigma_z$ is measured on one of the quantum systems of $\tilde{E}$, the MP is at a point either along $+z$ or $-z$ direction and this provides the source of preference to the observer. It has been established in the von Neumann projection postulate that whenever each individual member in the entangled state $|A_+\rangle$ or $|A_-\rangle$ is subjected to a measurement of an observable, e.g. $\sigma_z$, the entangled state splits into a mixture of states $|\phi_1\rangle = |\uparrow_z\rangle \otimes |h\rangle$ and $|\phi_2\rangle = |\downarrow_z\rangle \otimes |t\rangle$. The reduction of state vector occurs for individual members of the ensemble $\tilde{E}$. Measurement of $\sigma_z$ will not result in a mixture of $|A_+\rangle$ and $|A_-\rangle$ as described in Eq.(58). The preferred basis, in terms of which superpositions should be expanded is provided by the position of the MP, fixed by the observer. This point of view, already stated, is supported by Copenhagen Interpretation.

Apart from the preferred basis, there is another element [30] in the measurement problem which compares the pre-measurement entangled state of the (system + device) composite system:
$$|\psi_e\rangle = a|\psi_1\rangle|\phi_1\rangle + b|\psi_2\rangle|\phi_2\rangle$$
with the reduced density matrix
$$\rho_{red} = |a|^2|\psi_1\rangle\langle\psi_1| + |b|^2|\psi_2\rangle\langle\psi_2|. \tag{62}$$

These two different states must be compatible since no particular measurement result of the QS has been recorded. If an average of the observable $\hat{B}$ of the QS is measured, then

$$\langle B_e \rangle = \langle \psi_e | 1 \otimes B | \psi_e \rangle$$
$$= |a|^2 \langle \psi_1 | B | \psi_1 \rangle + |b|^2 \langle \psi_2 | B | \psi_2 \rangle \quad (63)$$

where I have utilized the orthogonality of $|\phi_1\rangle$ and $|\phi_2\rangle$ and the normalizability of $|\varphi\rangle$'s. The average value for the reduced density matrix, Eq.(62), is

$$\langle B_{red} \rangle = \text{trace}(\rho_{red} B)$$
$$= |a|^2 \langle \psi_1 | B | \psi_1 \rangle + |b|^2 \langle \psi_2 | B | \psi_2 \rangle \quad (64)$$

But these two averages, $\langle B_e \rangle$ and $\langle B_{red} \rangle$ are not equal because the reduced density matrix of Eq.(62) describes a mixture formed in measurement with no selection, i.e. the eigenfunctions of the quantum system are plane waves at the MP, and therefore the operator B in Eq.(64) is Hermitian. But the entangled state

$$\psi_e = a\theta_1 + b\theta_2$$

is in the pre-measurement state, which means that any of the probability wave peaks of $|\theta_1|^2$ or $|\theta_2|^2$ has not yet reached the MP. This means that any of the component eigenfunctions of the QS is not a plane wave. The observable $\hat{B}$ corresponding to eigenfunctions $\psi_1$ and $\psi_2$ (which are constituents of $\theta_1$ and $\theta_2$ respectively) is non-Hermitian. The average $\langle B_e \rangle$ will be a complex number while $\langle B_{red} \rangle$ will be a real number. In a forth-coming paper it will be shown that spin component eigenvalues are generally complex numbers, but at the MP they are real numbers. Hence there is a difference between the above two averages. This conclusion does not change even if we assign wave function to an individual quantum system, as we have already proved the von Neumann and Dirac projection postulates.

It is now the right place to consider a subtle difference between an actual measurement (where a permanent record is made) and the reduction of state vectors in the particular case when the QS is allowed to go past the MP. When a record is made the QS is stopped and absorbed. It is amazing that this is mathematically included in Quantum Mechanics ! We have found in section (2) that the total probability for a QS is (v/R). Normalization gives R=v. When the QS reaches the MP, we find from Eq.(43) that R=0 at the measurement event, which in turn says that the particle velocity v=0 at such an event. This unequivocally indicates stopping of the particle by the device particle at the MP. In case of a photographic plate, for example, the measurement constitutes absorption and a permanent record. In case where we let go the particle e.g. in a Stern-Gerlach device, the quanta of electromagnetic field (arising out of motion of the QS in a varying field) scatters the QS which finds its home in one of the sub-ensembles. Reduction of state vectors occur in both cases.

## 9. PROBABILITY FIELD CREATED BY A MOVING QUANTUM SYSTEM

To derive a probability field from the wave function of a QS, note that only two reference frames (RF) are at work in the measurement problem. The first is formed, say, with the uniformly moving QS as its origin. The second is formed by a particle of the measuring device at the MP. The unprimed quantities refer to the RF at the MP while the primed ones refer to the moving RF of the QS with respect to the MP. The Galilean transformations

$$x = vt' + s \quad \text{and} \quad t = t' \tag{65}$$

induces a transformation $\psi(x,t) = \psi'(s,t)\exp[if(x,t)]$ where s is the separation between the two RF's.
Assume that $f(x,t) = \theta_1(x) + \theta_2(t)$ is real. $\tag{66}$

Invariance of Schroedinger's equation requires [5]

$$\left.\begin{array}{c} \dfrac{\hbar}{m}\dfrac{\partial f}{\partial x} - v = 0, \\[6pt] \dfrac{\partial^2 f}{\partial x^2} = 0 \\[6pt] \text{and} \quad \dfrac{\hbar}{2m}\left(\dfrac{\partial f}{\partial x}\right)^2 - v\left(\dfrac{\partial f}{\partial x}\right) - \left(\dfrac{\partial f}{\partial t}\right) = 0 \end{array}\right\} \tag{67}$$

Substitution of Eq.(66) into Eq.(67) yields respectively

$$\frac{d\theta_1}{dx} = \frac{mv}{\hbar} = k = \frac{d\phi_1}{dx}, \text{ implying } \theta_1(x) = kx$$

$$\frac{d^2\theta_1}{dx^2} = \frac{dk}{dx} = 0$$

So, $k(x) = $ constant.

$$\frac{\hbar k^2}{2m} - kv = \frac{d\theta_2}{dt}; \text{ so, } \frac{d\theta_2}{dt} = k\left(\frac{v}{2} - v\right) = -\left(\frac{1}{2}kv\right).$$

The above equation yields $\theta_2 = -\left(\dfrac{1}{2}kvt\right) = -\omega t$, where $\dfrac{\omega}{k} = \dfrac{v}{2}$.

Eq.(66) is now rewritten as $f(x,t) = kx - \omega t$. In the above calculation we have neglected the constants of integration.

For a free particle

$$\psi(x,t) = \exp\left[\frac{R}{2}\left(t - \frac{x}{v}\right) + ikx - i\omega t\right]$$
$$= [\exp(if)]\psi'(s,t')$$
$$= [\exp(ikx - i\omega t)]\psi'(s,t')$$

so that $\quad \psi'(s,t') = \exp\left[\dfrac{1}{2}R\left(t - \dfrac{x}{v}\right)\right] = \exp\left[\dfrac{1}{2}(vt - x)\right],$ for $x \geq vt$

where R has been replaced by v. From Eq.(65)

$$s = x - vt' = x - vt$$

Therefore, $P'(s) = \exp(-s)$. (68)

Eq.(68) expresses the law of probability field created by moving QS around it: The probability of finding a QS at a point distant s from it is $e^{-s}$.

## 10. QUANTUM MECHANICS IN TEN DIMENSIONS AND TIME AS A NON-HERMITIAN OPERATOR

Eq.(7) states that the eigenposition of a free QS at any particular time is complex-valued. So far, as per convention, time has been treated as a parameter. To determine the trajectory of a QS in a potential we first note that the average position is

$$\langle s \rangle = \int_{s_0 = \int v dt}^{\infty} \psi^* s \psi ds = \int_{s_0}^{\infty} \psi \psi^* \left(\frac{s\psi}{\psi}\right) ds = \sum_i P_i(s) R_i(s).$$

From the above equation one immediately recognizes that

$$R_i(s) = \left(\frac{s\psi}{\psi}\right)$$

is the outcome of the i-th position measurement of the QS occurring with probability $P_i(s)$. Therefore the rule for obtaining any particular result (say jth) of an observable $\hat{A}$ in Quantum Mechanics is

$$R_i(A) = \left(\frac{A\psi}{\psi}\right). \qquad (68a)$$

For a QS in a potential the result of a position measurement on the QS is, according to Eq.(45),

$$x_c = \frac{s\psi}{\psi} = \left[\left(\int_0^{t_0} \frac{p}{m} dt\right)\psi\right]\frac{1}{\psi}$$

$$= \left(\int_0^{t_0} \frac{p\psi}{m} dt\right)\frac{1}{\psi}$$

$$= \left[v(x) + \left(\frac{i\hbar}{2m}\right)\left\{\frac{1}{v(x)}\frac{dv}{dx} + \frac{v_0}{v(x)}\right\}\right]t_0$$

The complex-valued position of the QS strongly indicates that it travels in complex space. In Sec.(2) it has been shown that momentum and energy of a QS are also complex. A straightforward calculation shows that the expectation or average values of position, momentum or energy carried out in the quantum domain $[x > \int v dt, or, x < \int v dt]$ are complex and therefore it requires a complex space-time as the background of quantum dynamics.

Complex eigenvalues of energy and momentum discussed in Sec.(2) for a free particle may be written as

$$E_c = E + iE' = \hbar\omega + \frac{i\hbar R}{2} = \frac{\hbar^2 k^2}{2m} - \frac{\hbar^2 R^2}{8mv^2} + \frac{i\hbar R}{2} \qquad (69)$$

and
$$p_c = p_x + ip'_x = \hbar k + \frac{i\hbar R}{2v}$$

The complex-valued energy of the free QS should be
$$\frac{p_c^2}{2m} = \frac{1}{2m}(p_x + ip'_x)^2$$
$$= \frac{1}{2m}\left[\frac{\hbar^2 k^2}{2m} - \frac{\hbar^2 R^2}{8mv^2} + \frac{i\hbar R}{2}\right]$$
$$= \hbar\omega + \frac{i\hbar R}{2}$$

which confirms Eq.(69). This energy equation requires R to have the dimension of frequency.

Complex numbers for position and momentum in the entire space (except at the MP) strongly suggest that quantum space-time is complex space-time. There is a trick here, because complex eigenvalues will always remain unmeasured by a device particle (at MP) since a QS will always be away from the MP for $x \neq vt$. Replacing x and $p_x$ by their complex counterparts $x_c$ and $p_c$ and rearranging $\psi(x,t)$ for a free QS, [Eq.(15)] we define

$$x_c = x + ix'$$

$$p_c = -i\hbar \frac{\partial}{\partial x_c}$$

$$t_c = t + it'$$

to obtain
$$\psi(x_c, t_c) = \exp[(1-i)(\omega t - kx)]$$
$$= \exp[ik(x+ix) - i\omega(t+it)]$$
$$= \exp[ikx_c - i\omega t_c]$$
$$\therefore \psi(x_c, t_c) = \exp[ikx_c - i\omega t_c] \qquad (70)$$

where $x' = x; t' = t$. Eq.(70) is the wave function of a free QS in complex space-time $(x_c, t_c)$, and for it

$$x_c = x + ix$$

and
$$t_c = t + it. \qquad (70a)$$

To find whether $\psi(x_c, t_c)$ satisfies the complex counterpart of Schroedinger's equation for a free QS

$$-\frac{\hbar^2}{2m}\frac{\partial^2 \psi_c}{\partial x_c^2} = i\hbar \frac{\partial \psi_c}{\partial t_c} \tag{71}$$

we insert $\psi(x_c, t_c) = \exp(ikx_c - i\omega t_c)$ in the above equation and find

$$E = \hbar\omega = \frac{\hbar^2 k^2}{2m}$$

and quite surprisingly the quantum potential term disappears in complex space-time. If, in general, $\psi(x_c, t_c)$ is found to satisfy a differential equation (71), then $\psi(x_c, t_c)$ must obviously be analytic. Using the operators $x_c$ and $p_c$ in complex space-time

$$\begin{aligned}
&(x_c p_c - p_c x_c)\psi(x_c, t_c) \\
&= x_c\left(-i\hbar \frac{\partial \psi}{\partial x_c}\right) + i\hbar\psi + i\hbar x_c \frac{\partial \psi}{\partial x_c} \\
&= i\hbar\psi(x_c, t_c); \text{ or, } x_c p_c - p_c x_c = i\hbar 1
\end{aligned} \tag{72}$$

which establishes the commutation relation of incompatible observables in quantum-mechanical complex space-time. Now I shall prove that $t_c$ is an operator or observable. Unlike the conventional tag of parameter it will be found that time is actually a non-Hermitian operator.

Note that the position coordinate of a free particle in classical physics is

$$x = vt = \frac{pt}{m}$$

In complex space-time the above relation is

$$x_c = \frac{p_c t_c}{m} \tag{73}$$

We quantize this equation and insert it in Eq.(72):

$$\frac{1}{m}(p_c t_c p_c - p_c p_c t_c)\psi_c = i\hbar\psi_c$$

where $\psi_c = \psi(x_c, t_c)$. If $t_c$ is not an operator, the above equation leads to

$$\psi_c = 0$$

for any $\psi_c$. $t_c$ is therefore an operator. Consequently, we symmetrize Eq.(73) as

$$x_c = \frac{1}{2m}(p_c t_c + t_c p_c)$$

and insert it in the operator equation (72) to obtain

$$\left[\frac{1}{2m}(p_c t_c + t_c p_c)p_c - p_c \frac{1}{2m}(p_c t_c + t_c p_c)\right]\psi_c = i\hbar\psi_c$$
$$\Rightarrow (t_c p_c p_c - p_c p_c t_c)\psi_c = 2im\hbar\psi_c$$

$$\Rightarrow \left[ t_c \left( \frac{p_c^2}{2m} \right) - \left( \frac{p_c^2}{2m} \right) t_c \right] \psi_c = i\hbar \psi_c$$

$$\Rightarrow t_c H_c - H_c t_c = i\hbar 1 \tag{74}$$

which is the much sought-after commutation relation. To prove that $t_c$ is a non-Hermitian operator the adjoint of Eq.(74) yields

$$\left( H_c^\dagger t_c^\dagger - t_c^\dagger H_c^\dagger \right) \psi_c^* = -i\hbar \psi_c^* \tag{75}$$

Now $H_c^\dagger \psi_c^* = (H_c \psi_c)^* = (E - iE') \psi_c^*$. Therefore, from Eq.(75)

$$H_c^\dagger t_c^\dagger \psi_c^* = t_c^\dagger H_c^\dagger \psi_c^* - i\hbar \psi_c^*$$

or

$$H_c^\dagger t_c^\dagger \psi_c^* = t_c^\dagger (E - iE') \psi_c^* - i\hbar \psi_c^* \tag{76}$$

Let $t_c$ be Hermitian, so that $t_c \psi_c = t_0 \psi_c$, where $t_0$ is the real eigenvalue of $t_c$.

Now, $(t_c \psi_c)^* = t_0 \psi_c^* \Rightarrow t_c^\dagger \psi_c^* = t_0 \psi_c^*$.

The above result is inserted into Eq.(76):

$$H_c^\dagger t_0 \psi_c^* = (E - iE') t_0 \psi_c^* - i\hbar \psi_c^*,$$

which implies

$$t_0 H_c^\dagger \psi_c^* = (E - iE') t_0 \psi_c^* - i\hbar \psi_c^*$$

or,

$$t_0 (E - iE') \psi_c^* = t_0 (E - iE') \psi_c^* - i\hbar \psi_c^*$$

i.e. $i\hbar \psi_c^* = 0$ for any $\psi_c^*$. $t_c$ is thus a non-Hermitian operator. Another proof for the operator nature of $t_c$ follows:

We consider the following well-known relation:

$$E = h\nu = \frac{h}{t} \tag{77}$$

The norm of $\hat{E} \psi_c$ is

$$\left\| i\hbar \frac{\partial \psi_c}{\partial t_c} \right\| = \sqrt{\left\langle i\hbar \frac{\partial \psi_c}{\partial t_c} \bigg| i\hbar \frac{\partial \psi_c}{\partial t_c} \right\rangle}$$

$$= \sqrt{\{i\hbar(-i\omega)\}^* i\hbar(-i\omega) \langle \psi_c | \psi_c \rangle} > b$$

where $\psi_c$ is given by Eq.(70) and b is positive number. So, $\hat{E}$ is invertible. The operator form of Eq.(77) now promotes time t to an operator:

$$t_c = h E_c^{-1}.$$

This operator nature of time elevates it on the same footing as any other spatial observable in quantum Mechanics. Especially $(x_c, y_c, z_c, t_c)$ forms a set of operators that is very much needed in a smooth formulation of Quantum Field theory [38]. If we add the complex spin variables $(\sigma + i\sigma')$ then description of a QS requires five complex-dimensional space-time or ten real-dimensional space-time. This might point towards a subtle relation to superstring theories [15]. The five imaginary dimensions of quantum space-time are obviously compactified, as they are not measurable, since a real space-time point created at the measurement event is the only window of observation . Peres [39] has remarked that "time is not a dynamical variable………commutator of t with any dynamical variable is always zero." But in complex space time it has been proved that the commutator $[t_c, H_c]$ is certainly not equal to zero.

## 11. HARTLE AND HAWKING`S NO- BOUNDARY PROPOSAL IN SINGULARITY-FREE QUANTUM COSMOLOGY AND INFORMATION LOSS PARADOX IN AN EVAPORATING BLACK-HOLE

Imaginary time is routinely used in quantum tunnelling and this concept is not new in Quantum Mechanics. J.B. Hartle and S.W. Hawking have proposed the notion of imaginary time in Quantum Cosmology to advance the conjecture that the boundary condition of the universe is that there is no boundary [51] . Since the universe at the big bang is a point particle it can be perfectly treated as a QS. At the big bang, real time is zero, and so there exists an option of using imaginary time of Quantum Mechanics in the background of complex space-time. In Fig [4] let the origin of time (i.e.the time of big bang) be at O. The quantum universe may be located at C at time $(t + it')$ in complex time. From this concealed site the universe may take any route to reach B and be observable in real time t during its journey if measurements are made.(There is a subtlety here that might trigger an explosive controversy : A device particle is needed for a measurement to occur ! It is a two body problem of Quantum Mechanics.)But the quantum universe can never reach the origin of time t = 0 at O. This is because space and time coordinates cannot have definite values. There will always be a spread. We have proved it for position and momentum coordinates while discussing the irreducible stochasticity of observables in Section (13) below. The same conclusion may be obtained from time-energy uncertainty relation in real space-time. The quantum universe will avoid the origin of time O and is free to climb up the imaginary time axis thus eluding the creation moment and may remain concealed there for an indefinite ("imaginary") time. So, that is the answer to the big question, "what happened before the big bang?" The universe continues to live in imaginary time without the breakdown of quantum laws. All points along the imaginary time axis are big bang points in real space-time: t = 0.

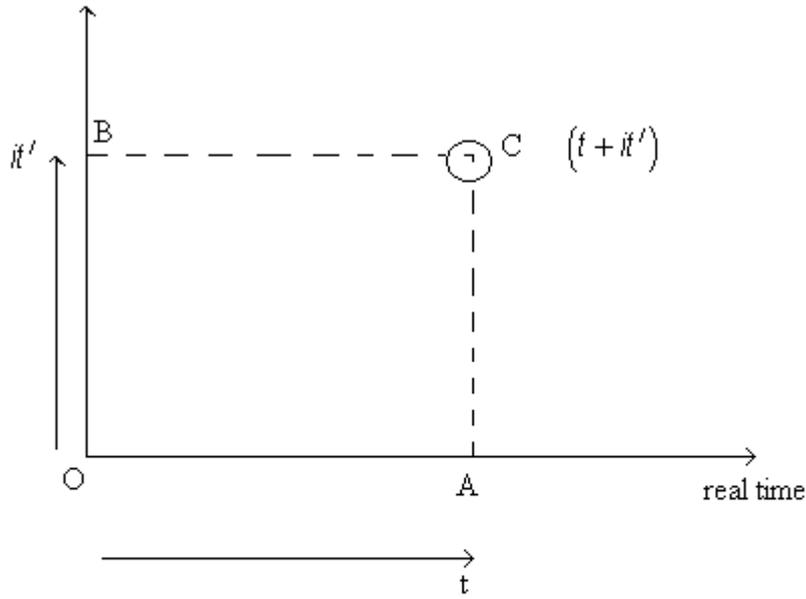

Figure(4): The initial temporal position of the quantum universe is at $C(= t + it')$, then it may travel along any path except AO to the point B where time is $it'$. All points along $it'$ are big bang points.

Since $P(x_c, t_c) = P_1(x_c) P_2(t_c)$ the integral $\int_{t+it'}^{it'} P_2(t_c) dt_c$ is independent of the path joining the complex times $t_c = t + it'$ and $t_c = it'$.

Singularity at the big bang, owing to zero volume, may be removed in Quantum Cosmology if we recall, from the discussion in Section (13) below, the irreducible stochasticity of position coordinates. The value of x, y, or z can never take up a definite value zero, because its uncertainty cannot be zero. A zero uncertainty is ruled out by Heisenberg`s position-momentum uncertainty relation because division by zero is not allowed by mathematics. On the other hand, multiplication by a zero uncertainty gives the absurd result that Planck`s constant h will be less than or equal to zero. The quantum universe will thus elude the singularity, and x,y and z will have stochastic values making room for a non-zero volume of the quantum universe. The non-zero values of x,y,and z imply that the quantum universe represented by a QS is not a point particle; but the size of the particle is finite albeit infinitesimal. Now we discuss Morera's theorem [18] for a specific reason: the quantum universe may have been created from nothing [19]. In Fig.[5] if $P_1(x_c)$ describes probability density of a QS at $x_c$ in the region $R'_1$ and similarly $P'_1(x_c)$ describes probability at $x_c$ in the region $R'_2$ then we describe $\tilde{P}(x_c)$ as

$$\tilde{P}(x_c) = P_1(x_c) \text{ for } x_c \text{ in } R'_1 \ ; \ P_1(x_c) = \exp\left[R\left(t_c - \frac{x_c}{v}\right)\right],$$

$$\text{or } = P'_1(x_c) \text{ for } x_c \text{ in } R'_2 \ ; \ P'_1(x_c) = \exp\left[R\left(\frac{x_c}{v} - t_c\right)\right]$$

so that $\tilde{P}(x_c)$ is analytic in the region $R' = R'_1 + R'_2$, and $P_1(x_c) = P'_1(x_c)$ on the line JKLM. The probabilities considered here are obviously complex probabilities. We have shown in section (13) below that Quantum Mechanics indulges their use formally. Morera's theorem of complex analysis states that

$$\underset{KSLTK}{\oint} \tilde{P}(x_c) dx_c = \underset{KSLK}{\oint} P_1(x_c) dx_c + \underset{KLTK}{\oint} P'_1(x_c) dx_c = 0$$

by cauchy's theorem [54]. $P'(x_c)$ is the analytic continuation of $P_1(x_c)$. $\tilde{P}(x_c)$ is now analytic in the entire space $R'_1 + R'_2$. This leads to an important result: if a quantum universe travels along any simple closed path then it is visible on real line at K, which provides the initial condition, and next at L on the real line at the measurement event. Hence the universe, having zero probability throughout its closed path, appears at K and

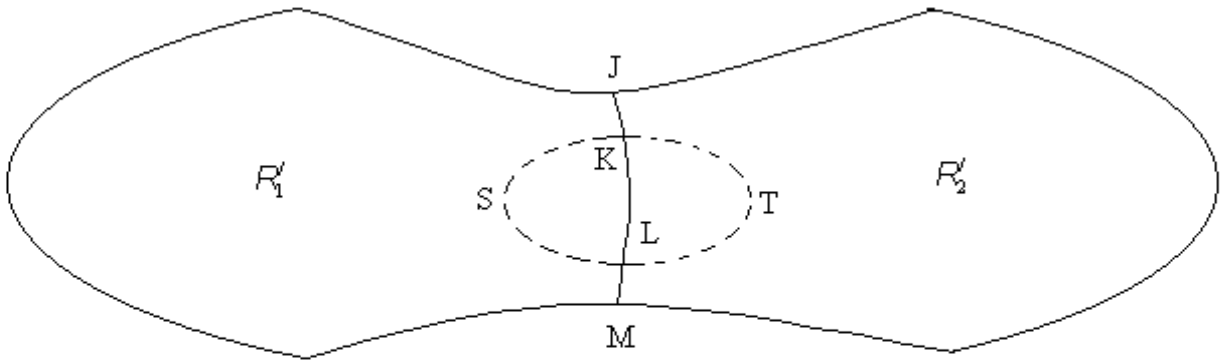

Figure(5): The figure shows that $\tilde{P}(x_c)$ is analytic in the region $R'_1 + R'_2$. The universe moving in a closed path appears at two points K and L on the real line while the net probability is zero.

L in complex space, but it traverses a non-closed path in real space boundary JKLM. This suggests that the quantum universe may have been created from nothing!

To have a feel of the trajectory of a QS in complex space-time, we note that the eigenposition of a free QS at a real time $t_0 < \frac{x}{v}$ is, from Eq. (7),

$$s = \left[ vt_0 + \frac{i\hbar R t_0}{2mv} \right].$$

If $t'_0 > \frac{x}{v} > t_0$, then from Eq.(14)

$$s' = \left[ vt'_0 - \frac{i\hbar R t'_0}{2mv} \right]$$

at $t = t'_0$. The position $s$ and $s'$ indicate that the QS has crossed the real x-axis during the time $(t'_0 - t_0)$.

We now return to the "information loss" paradox of an evaporating black hole. Information is not lost if there is no increase in entropy of an evaporating black-hole. Entropy does not increase in case of unitary evolution. It is an inescapable conclusion that if the time evolution operator is non-unitary, so that pure states evolve into mixed states, then entropy of an evaporating black-hole would increase resulting in information loss. A full theory of quantum gravity ought to be non-unitary [40]. We have shown that non-unitary evolution takes place between two measurements in Quantum Mechanics. Therefore there is an increase in entropy as is shown below.

At time t, if no measurement is made, the final entropy is

$$S_t = -K_B tr(\rho_t \ln \rho_t)$$

$$= -K_B \langle \ln \rho_t \rangle$$

$$= -K_B \left\langle \ln \left\{ \exp\left(-\frac{iHt}{\hbar}\right) \rho_0 \exp\left(\frac{iH^\dagger t}{\hbar}\right) \right\} \right\rangle$$

$$= -K_B \left\langle -\frac{iHt}{\hbar} + \ln \rho_0 + \frac{iH^\dagger t}{\hbar} \right\rangle$$

$$= -K_B tr(\rho_0 \ln \rho_0) - \frac{itK_B}{\hbar} \langle \psi_t | H^\dagger - H | \psi_t \rangle$$

$$= S_0 - \frac{itK_B}{\hbar} \left(-2iE' \langle \psi_t | \psi_t \rangle\right)$$

$$= 0 - \frac{2E' t K_B}{\hbar} (v/R)$$

$$= \left(-2\omega' t k_B v / R\right)$$

where $S_0$ is the entropy at t = 0. In the above we have used

$$\ln \langle \psi_0 | \psi_0 \rangle = 0,$$

because initial state was measured; and unnormalized $\langle \psi_t | \psi_t \rangle = v/R$, for, the state is unmeasured at t.

But, $(2E'/\hbar) = 2\omega' = R.$

Therefore, $\frac{dS_t}{dt} = -k_B v.$

So, during black hole evaporation, if no measurement is made at time t, there is a decrease in entropy from zero to a negative value i.e. negentropy. This implies that when left to itself, quantum entropy decreases during non-unitary time evolution [40a]. The consequent gain in information may thus be realized in a quantum-mechanical universe where, left to itself, it will progressively proceed towards order. But this information gain will be lost as soon as a measurement is made, which occurs at a classical point. The entropy then increases from the negative value to zero because the Hamiltonian now becomes Hermitian, and the difference $(H^+ - H)$ will be a zero operator. Thus, the second law of thermodynamics is switched on at a classical point, and information is indeed lost, as correctly inferred by Hawking. In a recent conference he has, however, retracted from this conclusion [21].

## 12. MEASUREMENT OF PROBABILITY DENSITY IN QUANTUM MECHANICS

The complex space-time is elusive because no measurement can be made in the quantum domain. But a crucial question may arise : How $P(x,t)$ can be measured at any point in the entire space except the MP?

The answer lies in identical measurements on individual quantum systems of an ensemble. We suppose there are 100 quantum systems in an ensemble. The probability of finding a QS at a fixed MP at x is, say, 40%. Then, out of 100 measurements only 40 particles are found at x because 40 probability wave peaks are found at x, and the rest 60 wave peaks are randomly distributed at points other than x.

## 13. UNCERTAINTY RELATION IN COMPLEX SPACE-TIME AND EXISTENCE OF NEGATIVE AND COMPLEX PROBABILITY IN QUANTUM MECHANICS

Real numbers being a subset of a set of complex numbers the Quantum space-time is essentially a complex space-time which includes the MP. To derive a position-momentum uncertainty relation in complex space-time we assume that if it is to satisfy Schroedinger equation, the wave function $\psi(x_c, t_c)$ must be analytic so that $\frac{\partial}{\partial x_c} \equiv \frac{\partial}{\partial x}$. The non-Hermitian observable $p_c$ may be written as

$$p_c = p_x + ip'_x$$

Where $p_x$ and $p'_x$ are Hermitian. The uncertainty in momentum of a QS is

$$(\Delta p_c)^2 = \langle p_c^2 \rangle - \langle p_c \rangle^2$$

$$= \langle p_x^2 \rangle - \langle p'^2_x \rangle + 2i\langle p_x \rangle\langle p'_x \rangle - \langle p_x \rangle^2 + \langle p'_x \rangle^2 - 2i\langle p_x \rangle\langle p'_x \rangle$$

$$= \left[\langle p_x^2 \rangle - \langle p_x \rangle^2\right] - \left[\langle p'^2_x \rangle - \langle p'_x \rangle^2\right]$$

$$= (\Delta p_x)^2 - (\Delta p'_x)^2$$

Therefore $\qquad (\Delta p_x)^2 = (\Delta p_c)^2 + (\Delta p'_x)^2 \qquad (78)$

Similarly, for the operator $x_c$,

$$(\Delta x)^2 = (\Delta x_c)^2 + (\Delta x')^2 \qquad (79)$$

In real space-time the Heisenberg position-momentum uncertainty relation yields

$$(\Delta x)^2(\Delta p_x)^2 = \left[(\Delta x_c)^2 + (\Delta x')^2\right]\left[(\Delta p_c)^2 + (\Delta p'_x)^2\right] \geq \frac{\hbar^2}{4} \tag{80}$$

If $\Delta x_c = 0$ and $\Delta p_c = 0$, the above inequality settles to

$$(\Delta x)^2(\Delta p_x)^2 = (\Delta x')^2(\Delta p'_x)^2 \geq \frac{\hbar^2}{4} \tag{81}$$

Also Eq.(74) leads to the following time-energy uncertainty relation:

$$(\Delta H)^2(\Delta t)^2 = \left[(\Delta H_c)^2 + (\Delta H')^2\right]\left[(\Delta t_c)^2 + (\Delta t')^2\right] \geq \frac{\hbar^2}{4}. \tag{82}$$

The above yields the well known relation

$$\Delta E \Delta t \geq \frac{\hbar}{2} \tag{82a}$$

where $\Delta t$ denotes the spread in the time of arrival t of the QS at the MP. Eq.(82a) corresponds to the commutation relation

$$Ht - tH = i\hbar \tag{82b}$$

An objection is often raised to the relation (82b): If t has a continuous spectrum then H must also have a similar spectrum, so that there could be no lower bound to energy, and no bound states with discrete energies. But a look at Figs.(1),(2) or (3) shows that time is measured from t = 0 to t = $\infty$. There is thus a lower bound to energy. For bound states, we recall from section (6) that in the Sturm-Liouville problem, a procedure was outlined for obtaining $v_n(x) = \hbar k_n(x)$. Since $t_n(x) = \int (dx/v_n)$ at the MP, the times of arrival will be discretized for discrete energies of a bound state. As both t and x are independent variables, the above relation implies that both will be constants at the MP. Thus discrete t discretizes x also. Therefore QM predicts discrete space-time for bound states. This is a strong support to loop quantum gravity which develops discrete geometry [21a]. A QS system may have definite complex-valued position and momentum as well as energy and time simultaneously, while the inequality in Eq.(81) merely describes that under these circumstances there are irreducible uncertainties in the real and imaginary parts of position and momentum of the QS. This is a reminder of the fact that indeterminacy enters into Quantum Mechanics through real-valued space-time. From an ontological point of view a QS is a particle moving deterministically in complex space-time. In real space-time it is a probability wave having no physical entity except at the MP where it is revealed as a particle. A QS shows indeterminism in its motion in real space-time through the stochasticity of its observables as its evident from the Heisenberg uncertainty relation:

$$\Delta x \Delta p_x \geq \frac{\hbar}{2}.$$

If $\Delta x$ or $\Delta p_x$ is zero then $\hbar \leq 0$, since division by zero is inadmissible. Therefore $\Delta x$ or $\Delta p_x$ is non-zero. Thus a QS cannot have definite position or momentum in real space time. The stochasticity of the MP at x helps divide the domain of states of the QS into two non-intersecting rigged Hilbert spaces $\Omega_+$ and $\Omega_-$ for the incoming and outgoing QS. One of the controversial issues of EPR paradox is whether a QS possesses a pre-measurement value of an observable. It has been proved that a QS has complex pre-measurement values of its observables. A deterministic dynamics as well as objective realities of a QS that Einstein firmly maintained, emerge when a QS travels in complex space-time. In spite of this, quantum nonlocality retains its "spooky" action at distance, which is demystified if it is realized that probability density wave is not a physical wave. Its mathematical entity with instantaneous influence cannot be considered as any physical signal. Hence it need not respect special relativity.

There has been a need for negative probabilities in different areas of quantum theory[60]. We are surprised to find that it follows easily from Quantum Mechanics. If x > vt, then $\pi_1(x)$ of a QS decreases with increase in x. Hence,

$$\frac{d\pi_1}{dx} = -|b| \tag{83}$$

where b is real.

But
$$d\pi_1(x) = P_1(x)dx$$

gives, from Eq.(83), $P_1(x) = -|b|$. Thus probability density is negative when a QS approaches the MP at x.

Complex probabilities may be allowed in Quantum Mechanics if we recognize that owing to analyticity,

$$\frac{\partial}{\partial x_c} \equiv \frac{\partial}{\partial x}.$$

If
$$\pi_1(x) = |\pi_1(x_c)|$$

then
$$\pi_1(x_c) = |\pi_1(x_c)|\exp[i\lambda(x_c)]$$

i.e.
$$\pi_1(x) = \pi_1(x_c)\exp(-i\lambda x_c) = \pi'(x_c)$$

so that
$$P_1(x) = \frac{d\pi_1}{dx} = \frac{\partial \pi'(x_c)}{\partial x} = \frac{\partial \pi'(x_c)}{\partial x_c}$$

which implies
$$d\pi'(x_c) = P_1(x)dx_c$$

i.e. elemental probability is complex [ 41.].

## 14. NORMALIZABILITY OF ALL WAVE FUNCTIONS AND AVERAGES OF SOME OBSERVABLES

We now seek to find a method to normalize any wave function. Consider the divergence of the integral

$$I = \int_{-\infty}^{\infty} \psi\psi^* dx$$

In case of a plane wave which represents a QS at the MP,
$$\psi(x,t) = \exp[ikx - i\omega t]$$

The divergence is usually cured by box or Dirac delta function normalization. Sometimes the rigged Hilbert space triplet which includes an extended space, is invoked to yield a convergent integral by confining k to only real values. An alternative normalization may be adopted for the plane wave. The indefinite integral

$$I = \int P(x,t)dx$$

is changed to a definite integral

$$I' = \int_0^\infty \psi\psi^* dx$$

For $x > vt$, one expects $\frac{\partial \pi(x,t)}{\partial x} = \pi_2(t)\frac{d\pi_1}{dx}$ to be negative, as $\pi_1(x)$ must decrease with increase in x. Therefore, $I' = -\int_{\pi=1}^{\pi=0} d\pi(x,t) = 1$.

Thus, without using explicit form of the wave function, it can be normalized by using the above prescription. Since most operators in Quantum Mechanics are normal operators these may be spectrally decomposed [61]. To find the relationship of probability distribution function $\pi(x,t)$ to a family of projection operators, I consider the operator

$$|x\rangle\langle x|.$$

Since

$$\langle |x\rangle\langle x|\rangle = \langle \psi |x\rangle\langle x|\psi \rangle = P(x,t) = \frac{\partial \pi}{\partial x}$$

position probability density is the average of the operator $|x\rangle\langle x|$.

Also, $$\pi(x,t) = \int d\pi(x,t) = \int \langle \psi |x\rangle\langle x|\psi \rangle dx = \langle \psi |\left(\int |x\rangle\langle x|dx\right)|\psi \rangle \qquad (83)$$

and so $\pi(x,t)$ is the average of the operator $\left(\int |x\rangle\langle x|dx\right)$ over the interval $[0,x]$. Converting it to a definite integral

$$[\pi(x,t)]_0^1 = 1 = \langle \psi |1|\psi \rangle$$

where the closure relation has been used. Total probability is thus the average of the identity operator over the relevant space.

**CONCLUSION**

To summarize, we have answered the questions: ``when``, ``where`` and ``why`` the reduction of state vectors occurs.The measurement problem is quite intractable if one insists on using only the Schroedinger unitary evolution . The source of the problem may be traced back to the use of stationary state solutions of Schroedinger equation. If we treat the spatial and temporal parts of the wave function as complex functions themselves, we obtain dynamical probability waves. These in turn generate the scope using non-Hermitian operators which become Hermitian at the field point or measurement point (MP) .The quantum leap (jump!) of Hermitian to a non-Hermitian Hamiltonian during evolution time proves that a QS evolves non-unitarily between successive measurements. A pure entangled state can easily split itself into a mixed state when an abrupt change from non-unitarity to unitarity occurs at the MP..Unitarity is restored only at the measurement event.We show the loss of coherence terms in the density matrix of a pure state of a composite system,when only one of the component density matrices survives.This is due to the presence of the particle at the peak of the corresponding probability wave. To prove this for an individual measurement process we derive both Dirac and von Neumann projection postulates from the standard formalism.By utilizing the quantum counterpart of Hamilton-Jacobi equation we prove that the field point or measurement point is actually a classical space-time point for a momentary period of time.Two new wave functions for the free and non-free states have been derived. The eigenpositions of these particles forced us to consider complex space-time which is the long-awaited quantum space-time. Two theorems prove that time is not a parameter but a non-Hermitian normal operator. The derivations of both position-momentum and energy-time uncertainty relations in complex space-time, whose special cases are Heisenberg uncertainty relations unequivocally shows that a QS travels with simultaneous precise values of both position and momentum. It was recognized that in the context of wave-particle duality, a QS travels as a probability wave when left to itself ( i.e. no measurement is made on it) ;but at a measurement event it reveals itself as a particle in real space-time.We now turn to quantum cosmology and find that Hartle-Hawking`s no-boundary proposal can ensure a smooth passage of quantum universe into imaginary time, where it may remain concealed for an indefinite (imaginary) time. Using stochastic position observables we find that the big bang singularity may be removed. A decrease in entropy of an evaporating black hole has been computed with the help of non-unitarity ;but at the measurement point it suddenly gains some entropy.Therefore,there is a genuine loss of information in an irreversible manner.The demystified motion of a QS is as follows: the system travels as a particle in complex space-time as long as there is no measurement. As a vestige of the QS a probability wave travels in probability space during this time. When the peak of the probability wave carrying the particle reaches the measurement point, a whole particle pops up from complex space-time into real space-time. The most crucial point is that Quantum Mechanics is a universal theory.We live in complex space-time. It is indeed true that there is more to Quantum Mechanics than meets the eye.

I sincerely thank K. Ghosh, K. Goswami, C .Patra, D. Home, B.B Bal, P.B.Pal, A.Majumdar, M. Gupta, A.K.Nag, A. Roypradhan , H.Chatterjee, S. Mukherjee and J. Chatterjee for valuable help.